\documentclass[conference]{IEEEtran}
\ifCLASSINFOpdf
\else
\fi
\usepackage{multirow}
\usepackage[pdftex]{graphicx,color}
\usepackage[square,comma,sort&compress,numbers]{natbib}
\usepackage{amsmath}
\usepackage{amssymb}
\newtheorem{lemma}{Lemma}
\DeclareMathSizes{10}{8}{7}{6}
\IEEEaftertitletext{\vspace{-2\baselineskip}}
\begin{document}
\title{Design of Raptor Codes in the Low SNR Regime with Applications in Quantum Key Distribution}
\author{\IEEEauthorblockN{Mahyar Shirvanimoghaddam\IEEEauthorrefmark{1}, Sarah J. Johnson\IEEEauthorrefmark{1}, and Andrew M. Lance\IEEEauthorrefmark{2}}
\IEEEauthorblockA{\IEEEauthorrefmark{1}School of Electrical Engineering and Computer Science, The University of Newcastle, NSW, Australia}
\IEEEauthorblockA{\IEEEauthorrefmark{2}Quintessence Labs Pty. Ltd., Canberra, ACT, Australia\\
Emails:\{mahyar.shirvanimoghaddam, sarah.johnson\}@newcastle.edu.au, AL@quintessencelabs.com}}
\maketitle

\begin{abstract}
The focus of this work is on the design of Raptor codes for continuous variable Quantum key distribution (CV-QKD) systems. We design a highly efficient Raptor code for very low signal to noise ratios (SNRs), which enables CV-QKD systems to operate over long distances with a significantly higher secret key rate compared to conventional fixed rate codes. The degree distribution design of Raptor codes in the low SNR regime is formulated as a linear program, where a set of optimized degree distributions are also obtained through linear programming. Simulation results show that the designed code achieves efficiencies higher than 94\% for SNRs as low as -20 dB and -30 dB. We further propose a new error reconciliation protocol for CV-QKD systems by using Raptor codes and show that it can achieve higher secret key rates over long distances compared to existing protocols.
\end{abstract}

\IEEEpeerreviewmaketitle
\section{Introduction}
Quantum key distribution (QKD) enables two spatially-separated parties to share a secure random key, which is the first practical application of quantum physics. In QKD systems, the secret key is established through exchanging quantum states, even in the presence of an eavesdropper, with the help of a classical auxiliary authenticated communication channel. Two major approaches to QKD include discrete variable QKD (DV-QKD), which uses single-photon or weak coherent states with a single photon detector, and continuous variable QKD (CV-QKD), where coherent states with homodyne detectors are used. These approaches have shown to be information-theoretic secure \cite{PhysRevLett}.  Due to recent technological advancement in quantum physics, CV-QKD is shown  to effectively overcome some limitations of DV-QKD systems, which are mainly due to the speed and efficiency of single photon detectors.  This is achieved by encoding information on continuous variables such as the quadrature of coherent states.

In conventional standard CV-QKD protocols, Gaussian modulated coherent states are prepared and measured by a homodyne or hetrodyne detector \cite{PhysRevA}. Although, this is optimal in terms of the theoretical security \cite{PhysRevLett}, the efficiency of current reconciliation protocols for Gaussian variables dramatically drops in the low SNR regime. Non-Gaussian modulations, either discrete \cite{PhysRevAD11} or continuous \cite{PhysRevAC11}, were then proposed to overcome this problem as they theoretically increase the achievable secure distance of CV-QKD systems. They are also compatible with high-performance error correction codes, which enables them to efficiently extract the information available in their raw data. In this regard, \cite{PhysRevA} proposed a highly efficient error correction code, which can be combined with a multidimensional reconciliation scheme, where a secret key exchange was enabled for a CV-QKD system with Gaussian modulation in the low SNR regime.  In fact,  to achieve high key generation rates and longer transmission distances, the reconciliation step on the classical communication channel has to correct key errors close to the theoretical limits to minimize the amount of information leaked to a potential attacker. Therefore, very low rate capacity approaching error control codes are necessary in the error reconciliation stage of CV-QKD systems.

Iterative error control codes have the potential to substantially improve the efficiency of CV-QKD systems. For instance, repeat accumulate (RA) codes have been investigated to use in CV-QKD systems in \cite{Sarah_J_RA_QKD}, where the codeword length is 64800 and the lowest code rate is 1/20. Although, the same design approach along with puncturing and extension techniques can be used to design lower rate RA codes, these codes have shown to have poor efficiencies in very low SNRs, which limit their applications in long distance CV-QKD systems. Moreover, multi-edge type LDPC (MET-LDPC) codes were shown in \cite{PhysRevA} to theoretically approach high efficiencies in low SNRs. However, existing very low rate codes, such as RA and ME-LDPC codes, have a very poor word error rate performance and so will only produce good efficiencies for theoretical threshold results or for finite-length results where the allowed word error rate is very high.

To overcome this problem and enable CV-QKD systems to operate over very long distances, we propose to use rateless codes in the error reconciliation step. Binary rateless codes are special types of code on graphs, which can potentially generate an unlimited number of coded symbols. Thanks to the interesting properties of graph codes, Luby proposed the first practical realization of rateless codes, namely the Luby Transform (LT) codes \cite{Luby}. These codes have very simple encoding and decoding processes and can approach the capacity of binary erasure channels (BECs) with an unknown erasure rate. A more practical extension of these codes, namely Raptor codes, can be obtained by precoding the entire data using a high rate LDPC code and then use an LT code to generate an unlimited number of coded symbols. The encoding and decoding of these codes are linear in terms of the message length; thus practical for several applications with large data transmission.

Raptor codes were studied for additive white Gaussian noise (AWGN) channels in \cite{RaptorBSC}, where a systematic framework was proposed to find the optimal degree distribution in all signal to noise ratios (SNRs). More specifically, a linear program was proposed to find the optimal degree distribution for a capacity approaching code. The authors in \cite{Raptor_BIAWGN} have studied the feasibility of this linear program and shown that it is only feasible for SNR larger than a certain value, which depends on the average variable node degree. However, as mentioned in \cite{Raptor_WideSNR,Raptor_BIAWGN}, the linear program fails to provide good degree distributions in very low SNRs (below -10 dB).

In this paper, we study the design of Raptor code in the low SNR regime, where the existing linear program is approximated through several mathematical tools and a more practical linear program is then proposed. A set of optimal degree distributions are then derived, which show excellent performance in very low SNRs. This motivates us to use these codes in CV-QKD systems, where highly efficient very low rate codes can significantly improve the secret key rate and operational distances. For this aim, we slightly modify the conventional error reconciliation protocol to adapt with the rateless nature of Raptor codes. Numerical results show that the proposed scheme provides near optimal secret key rate and outperforms existing fixed-rate protocols in terms of the secret key rate in different operational distances.


The rest of the paper is organized as follows. Section II presents an overview on Raptor codes. In Section III, we present the general framework to design the degree distribution of Raptor codes and propose a linear program for the design of Raptor codes in the low SNR regime. We then show in Section IV, how Raptor codes can be used in CV-QKD systems, where we propose a new error reconciliation protocol for CV-QKD systems. Numerical results are presented in Section V. Finally, Section VI concludes the paper.
\section{An Overview on Raptor Codes}
We first provide a brief overview on the two most practical rateless codes, referred to as LT and Raptor codes. These codes are rateless in the sense that they can produce an infinite number of coded symbols until the destination receives sufficient information to perform successful decoding. Each LT code is characterized by a message length $k$ and a degree distribution function. The encoding process of LT codes contains two important steps. First, an integer $d$, called degree, is obtained from a predefined probability distribution function, called a degree distribution. Second, $d$ distinct information symbols are uniformly selected at random and then XORed to produce one coded symbol. The encoding process will be terminated when the sender receives an acknowledgement from the destination or a pre-determined number of coded symbols are sent.

\begin{figure}
  \centering
  \includegraphics[scale=0.21]{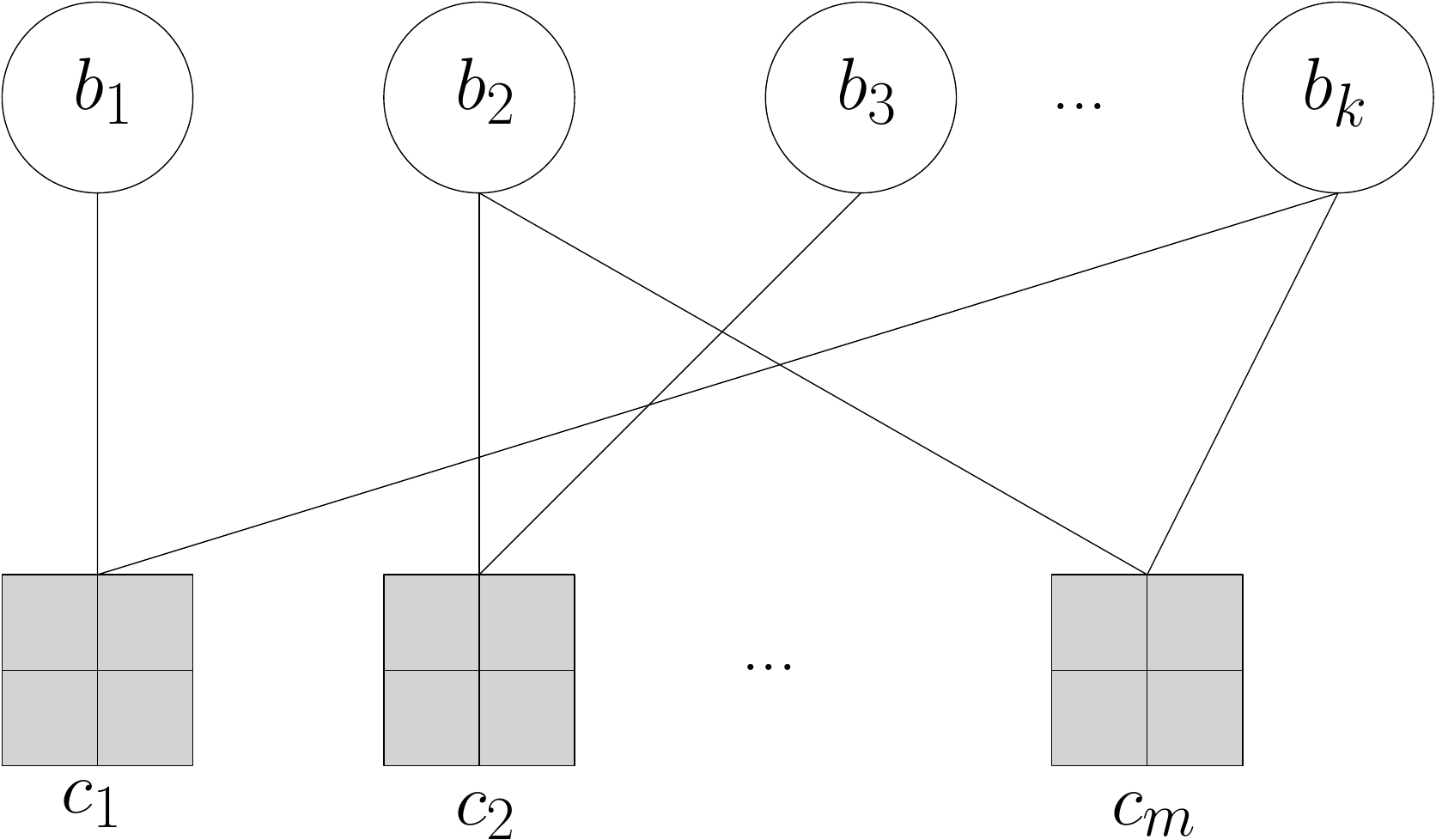}
  \caption{The factor graph of an LT code truncated at code length $m$.}
  \label{Fig1}
\end{figure}

Let $\Omega_d$ denote the probability that the degree is $d$. Then, the degree distribution function can be represented in a polynomial form as follow:
\begin{align}
\Omega(x)=\sum_{d=1}^{D}\Omega_dx^d,
\label{Eq1}
\end{align}
where $D$ is the maximum code degree. The degree distribution with respect to edges is defined by $\omega(x)=\Omega'(x)/\Omega'(1)$ \cite{Raptor}, where $\Omega'(x)$ is the derivative of $\Omega(x)$ with respect to $x$ and $\beta :=\Omega'(1)$ is the average output node degree. The degree distribution can be defined from input node and edge perspectives, and it has been shown in \cite{Raptor} that it is asymptotically $e^{\alpha(x-1)}$, when the message length goes to infinity, where $\alpha$ is the average input node degree. The LT encoding process can be represented by a bipartite graph, when information and coded symbols are mapped to variable and check nodes, respectively. Fig. \ref{Fig1} shows the bipartite graph of an LT code truncated at code length $m$.

Since information symbols in the LT encoding process are selected uniformly at random, it is probable that some information symbols are not selected by any coded symbol. Therefore, these information symbols cannot be decoded at the receiver. To overcome this problem, Shokrollahi [2] proposed to use a high rate precoder before LT encoding. The entire code is then called a Raptor code, which is characterized by a precoder $\mathrm{C}$ of a fixed rate, a message length $k$, and a degree distribution function. In other words, a Raptor code is a simple concatenation of a high rate code (usually an LDPC code due to its simplicity) and an LT code.  A sum product algorithm (SPA) is usually used for the decoding of Raptor codes, where log-likelihood ratios (LLRs) are passed as messages along edges from variable to check nodes and vice versa in an iterative manner. More details of this decoder can be found in \cite{RaptorBSC}.


The realized rate of a Raptor code is defined as the ratio of the message length, $k$, and the average number of coded symbols required for the successful decoding of the entire message at the destination. In a practical system, feedback is used to adjust the total number of coded symbols sent for each message. Let $n(\gamma)$ denote the number of coded symbols required for the receiver to successfully decode a message of length $k$ using a Raptor code at SNR $\gamma$. The realized rate of the Raptor code is then given by:
\begin{align}
R_r(\gamma)={k}/{\mathbb{E}[n(\gamma)]},
\label{Eq2}
\end{align}
where $\mathbb{E}$ is the expectation operator. The rate efficiency of the code at SNR $\gamma$, denoted by $\eta(\gamma)$,  is then defined as the ratio of the realized rate and channel capacity $C(\gamma)$, i.e.,
\begin{align}
\eta(\gamma)={R_r(\gamma)}/{C(\gamma)}.
\label{Eq3}
\end{align}

\section{Raptor Codes in the Low SNR Regime}
\subsection{Degree Distribution Optimization}
In \cite{RaptorBSC}, a framework to optimize the degree distribution of a Raptor code for a given BI-AWGN channel has been proposed. This framework, referred to as the mean-LLR-Exit chart approach, was proposed based on two assumptions. The first is the cycle-free assumption, which means that the factor graph representation of Raptor codes is locally cycle-free, which is well justified when the graph is large and sparsely connected. This assumption enables us to assume that all incoming SPA messages arriving at a given node are statistically independent. The second assumption is that the probability density of a message passed from an input symbol to an output node along a randomly chosen edge in the graph is a mixture of symmetric Gaussian distributions. This can be satisfied when the degree of each input node is relatively large, then according to the Central Limit Theorem the sum of incoming messages to an input node follows a Gaussian distribution. The mixture of Gaussian model is then due to the irregular degrees of input nodes \cite{Raptor_BIAWGN}.

Using the SPA decoder and the assumption that the messages are mutually independent symmetric Gaussian random variables, the mean value of the LLR message from a check node of degree $d$ to a neighboring variable node can be calculated as follows \cite{RaptorBSC}:
\begin{align}
f_d(\mu):=2\mathbb{E}\left[\tanh^{-1}\left(\tanh\left(\frac{Z}{2}\right)\prod_{q=1}^{d-1}\tanh\left(\frac{X_q}{2}\right)\right)\right],
\label{ElemExit}
\end{align}
where $Z$ is the channel LLR, which follows a symmetric Gaussian distribution with mean $2\gamma$ and variance $4\gamma$, and $X_q$ is the incoming LLR message from a neighboring variable node, which is also modeled by a symmetric Gaussian distribution with mean $\mu$ and variance $2\mu$, for $q=1,\cdots, d-1$. As the outgoing message from a variable node to an adjacent check node in the SPA decoding is the sum of incoming LLR messages from the adjacent check nodes, the average value of the outgoing message from a variable node can be calculated by $\alpha\sum_{d=1}^D\omega_d f_d(\mu)$. This arises from the fact that the average number of edges connected to each variable node is $\alpha$ and the probability that a randomly chosen edge is connected to a check node of degree $d$ is $\omega_d$ \cite{RaptorBSC}.

Accordingly, \cite{RaptorBSC} proposed the following linear program to maximize the design rate of the LT code, defined as $R_{\mathrm{design}}:=\beta/\alpha$, for given $\alpha$, $D$, and the target maximal message mean $\mu_o$:
 \begin{align}
 \label{optorig}
\text{minimize}&~~\textstyle\alpha\sum_{d=1}^{D}\omega_d/d\\
\nonumber \text{s.t.}~& (i). ~\textstyle\alpha\sum_{d=1}^D\omega_{d}f_d\left(\mu_j\right)>\mu_j, ~~\forall j=1,\cdots,N,\\
\nonumber~&(ii).~\textstyle\sum_{d=1}^{D}\omega_d=1,\\
\nonumber~&(iii).~\textstyle\omega_d\ge 0, ~~\forall d=1,\cdots, D,
 \end{align}
where $\{\mu_j|j=1,\cdots,N\}$ is a set of equally spaced values in range $(0,\mu_o]$ and $\mu_N=\mu_o$. The constraint ($i$) in the above linear program is to make sure that the average LLR message is increased after each iteration of the SPA decoding. It is important to note that $\mu_o$, the mean bit LLR at the output of the LT decoder, is chosen to be large enough to ensure that the decoding of outer code $\mathrm{C}$ is successful.

The authors in \cite{Raptor_BIAWGN} provided a detailed study of the feasibility of the linear program (\ref{optorig}) based on parameters $\mu_o$ and $\alpha$. More specifically, they showed that for SNRs higher than $\text{SNR}_{\mathrm{low}}=\mu_o/2\alpha$, the linear program is feasible and thus has a solution. The following lemma shows that for any SNR, the linear program (\ref{optorig}) is always feasible in the asymptotic case, where the maximum degree goes to infinity.
\begin{lemma}
\label{feasibility}
Linear program (\ref{optorig}) is feasible for every SNR, when the average input node degree goes to infinity.
\end{lemma}
\begin{IEEEproof}
The proof of this lemma follows directly from the proof of Lemma 2 and Theorem 1 in \cite{Raptor_BIAWGN}. More specifically, $f_d(\mu)$ is shown to be increasing with $\mu$ for $d>1$, and also it decreases with $d$ for any $\mu$ and SNR. Therefore, we have:
\begin{align}
\nonumber f_d(\mu)\le f_1(\mu)=2\mathrm{SNR}.
\end{align}
Therefore, for constraint $(i)$ in (\ref{optorig}), we have:
\begin{align}
\nonumber \alpha\sum_{d=1}^{D}\omega_df_d(\mu_i)\le \alpha\sum_{d=1}^{D}\omega_d f_{1}(\mu_i)=\alpha f_{1}(\mu_i)=2\alpha\mathrm{SNR},
\end{align}
which directly follows from (\ref{ElemExit}) by substituting $d$ with 1 and the fact that $\mathbb{E}[Z]=2\gamma$. Thus, we have $\mu_i<2\alpha\mathrm{SNR}$. For the maximum value of $\mu_i$, i.e., $\mu_o$, we also have $\mu_o\le 2\alpha\mathrm{SNR}$. Now suppose that $\mathrm{SNR}>\mu_o/2\alpha$, then it is clear that $\omega(x)=1$ satisfies all constraints in (\ref{optorig}), thus the optimization is feasible.  In an asymptotic case, when the maximum degree is very large, $\alpha$ is potentially infinite, thus the optimization problem is feasible for $\mathrm{SNR}>\lim_{\alpha\to \infty}\mu_o/2\alpha = 0$. This completes the proof.
\end{IEEEproof}
\subsection{Degree Optimization in the Low SNR Regime}
Let us have a closer look at function $f_d(\mu)$ defined in (\ref{ElemExit}), when the channel SNR, $\gamma$, goes to zero. We define the function $h_d(Z,\textbf{X}_d)$ as follows:
\begin{align}
h_d(Z,\textbf{X}_d)=2\tanh^{-1}\left(\tanh\left(\frac{Z}{2}\right)\prod_{q=1}^{d-1}\tanh\left(\frac{X_q}{2}\right)\right),
\end{align}
where $\textbf{X}_d\triangleq(X_1,...,X_{d-1})$. Then it is clear that $f_d(\mu)=\mathbb{E}[h(Z,\textbf{X}_d)]$.  It can be easily verified that the even-order derivatives of $h_d(Z)$ at $Z=0$ are zero; therefore, the Maclaurin series of $h_d(Z,\textbf{X}_d)$ can be shown as follows:
\begin{align}
h_d(Z,\textbf{X}_d)=ZP_d+\frac{Z^3}{12}\left(P_d^3-P_d\right)+\mathcal{O}(Z^5)
\end{align}
where $P_d\triangleq\prod_{q=1}^{d-1}\tanh\left(\frac{X_q}{2}\right)$. Then, $f_d(\mu)$ can be approximated as follows:
\begin{align}
\label{2ndappfd}
f_d(\mu)&=\mathbb{E}[h_d(Z,\textbf{X}_d)]\approx\mathbb{E}\left[ZP_d+\frac{Z^3}{12}\left(P_d^3-P_d\right)\right].
\end{align}
Let $M_{2n-1}(x,2x)$ denote the $(2n-1)^{th}$ moment of a Gaussian distribution with mean $x$ and variance $2x$, which can be found as follows \cite{Mathhandbook}:
\begin{align}
M_{2n-1}(x,2x)=\sum_{j=0}^{n-1}\dbinom{2n-1}{2j}(2j-1)!! 2^{j}x^{2n-j-1},
\label{highmom}
\end{align}
and the double factorial for odd numbers is defined as $(2j-1)!!=(2j-1)(2j-3)\cdots5.3.1$. 
It is easy to verify that the minimum degree of $M_{2n-1}(x,2x)$ is $n$. Therefore, for $\gamma$ very close to zero, $\mathbb{E}(Z^{2n-1})$ is of $\mathcal{O}(\gamma^n)$. Then (\ref{2ndappfd}) can be further simplified as follows:
\begin{align}
\nonumber f_d(\mu)&\approx 2\gamma\prod_{q=1}^{d-1}\mathbb{E}\left[\tanh\left(\frac{X_q}{2}\right)\right]+\mathcal{O}(\gamma^2),
\approx 2\gamma\prod_{q=1}^{d-1}\mathbb{E}\left[\tanh\left(\frac{X_q}{2}\right)\right]
\end{align}
which follows from the fact that $Z$ follows a symmetric Gaussian distribution with mean $2\gamma$, $X_q$ and $Z$ are mutually independent random variables for $q=1, \cdots, d$, and $\gamma$ goes to zero. we can define function $\varphi(\mu)$ for $\mu\ge0$ as follows:
\begin{align}
\varphi(\mu)=\mathbb{E}\left[\tanh\left(\frac{X}{2}\right)\right]
=\frac{1}{\sqrt{4\pi \mu}}\int_{-\infty}^{\infty}\tanh\left(\frac{u}{2}\right)\text{e}^{-\frac{(u-\mu)^2}{4\mu}}du.
\label{phifunc}
\end{align}
Thus, $f_d(\mu)$ can be further simplified as follows:
\begin{align}
f_d(\mu)\approx2\gamma\left[\varphi(\mu)\right]^{d-1}.
\end{align}
The left hand side of the condition ($i$) in (\ref{optorig}) can then be rewritten as follows:
\begin{align}
\alpha\sum_{d}\omega_{d}f_d\left(\mu\right)\approx2\gamma\alpha\sum_{d}\omega_{d}\left[\varphi(\mu)\right]^{d-1}.
\label{Eq11}
\end{align}
It is important to note that in the low SNR regime, by using the well-know Maclaurin series of function $\ln(1+\gamma)$, the BI-AWGN channel capacity can be approximated as follows:
\begin{align}
C_b(\gamma)=\frac{1}{2}\log_2(1+\gamma)=\frac{1}{2\ln(2)}\left(\gamma+\mathcal{O}(\gamma^2)\right).
\label{appcap}
\end{align}
Therefore, for a Raptor code with rate efficiency $\eta$ at very low SNR $\gamma$, we have:
\begin{align}
\beta=\eta \alpha C_b(\gamma)=\frac{\eta\alpha}{2\ln(2)}\left(\gamma+\mathcal{O}(\gamma^2)\right).
\label{Eq22}
\end{align}
Accordingly, by using (\ref{Eq11}) and (\ref{Eq22}) we have:
\begin{align}
\frac{\eta\alpha\sum_{d}\omega_{d}f_d\left(\mu\right)}{4\ln(2)}&\approx\beta\sum_{d}\omega_{d}\left[\varphi(\mu)\right]^{d-1}
=\sum_dd\Omega_d[\varphi(\mu)]^{d-1}.
\label{finalcondi}
\end{align}
We can then reformulate the optimization problem for a given maximum degree $D$ as follows, when $\gamma$ goes to zero:
\begin{align}
 \label{optlowsnr1}
\text{maximize}&~~\textstyle\sum_{d=1}^D d\Omega_d\\
\nonumber\text{s.t.}~& (i). ~\textstyle\sum_{d=1}^Dd\Omega_{d}[\varphi(\mu_j)]^{d-1}>\frac{\eta(\mu_j+\epsilon)}{4\ln(2)}, \\
\nonumber~&(ii).~\textstyle\sum_{d=1}^{D}\Omega_d=1,\\
\nonumber~&(iii).~\textstyle\Omega_d\ge 0, ~~\forall d=1,..., D,
\end{align}
where $j=1,\cdots,N$ and we introduced $\epsilon>0$ in condition ($i$) to have a non-zero $\Omega_1$. This is because $\varphi(0)=0$, and when $\epsilon=0$, the optimal value of $\Omega_1$ would be zero. This means that in the first iteration of the SPA decoding, the LLR values cannot be updated as the average LLR is not increasing. Therefore, a nonzero $\epsilon$ is necessary to guarantee that the decoder can proceed and converge to a nonzero value. The above optimization is linear for a given $\eta$, $\mu_o$ and $D$. However, we consider a joint optimization problem, where instead of supplying a predetermined $\eta$, the above linear program is converted to a general non-linear optimization problem with the same objective and constraints as in (\ref{optlowsnr1}), but optimizing jointly over $(\eta,\Omega(x))$. For this aim, an optimal $\eta$ can be found by searching over a relatively large range of discretized values of $\eta$, where for each $\eta$ the optimization is converted to a linear program. 
\subsection{Rate Efficiency Results}
\begin{table}[t]
\caption{Optimized degree distributions and their corresponding maximum efficiencies $\eta$ and average degree, $\beta$, when $\epsilon=0.01$.}
\scriptsize
\centering
\begin{tabular}{|c||c|c|c|c|c|}
 \hline
$D$ &\multicolumn{2}{|c|}{100} &\multicolumn{2}{|c|}{300}\\
\hline
$\mu_o$&30&40 &30& 40\\
\hline
\hline
$\Omega_1$&0.0035&0.0034 &0.0034&0.0035\\
\hline
$\Omega_2$&0.3493&0.3397 &0.3574&0.3538\\
\hline
$\Omega_3$&0.2314&0.2095&0.2377&0.2338\\
\hline
$\Omega_4$&0.0624&0.1256&0.0651&0.0737\\
\hline
$\Omega_5$&0.1115&-&0.1036&0.0755\\
\hline
$\Omega_6$&-&-&-&0.0262\\
\hline
$\Omega_7$&-&0.1462&0.0316&0.0608\\
\hline
$\Omega_8$&0.0436&-&0.0622&-\\
\hline
$\Omega_9$&0.0696&-&-&-\\
\hline
$\Omega_{11}$&-&-&-&0.0493\\
\hline
$\Omega_{12}$&-&-&-&0.0255\\
\hline
$\Omega_{13}$&-&-&0.0382&-\\
\hline
$\Omega_{14}$&-&-&0.0242&-\\
\hline
$\Omega_{17}$&-&0.0337& -& - \\
\hline
$\Omega_{18}$&-&0.0495&-&-\\
\hline
$\Omega_{20}$&0.0286&-&- & - \\
\hline
$\Omega_{21}$&0.0401&-&- &0.0002\\
\hline
$\Omega_{23}$&-&-&- &0.0454\\
\hline
$\Omega_{26}$&-&- &0.0096&-\\
\hline
$\Omega_{27}$&-&- &0.0292&-\\
\hline
$\Omega_{57}$&-&- &-&0.0072\\
\hline
$\Omega_{58}$&-&- &-&0.0180\\
\hline
$\Omega_{66}$&-&-&0.0179&-\\
\hline
$\Omega_{67}$&-&- &0.0039&-\\
\hline
$\Omega_{100}$&0.0599&0.0924&- &-\\
\hline
$\Omega_{300}$&-&-&0.0158&0.0272\\
\hline
\hline
$\beta$&10.5821&13.5444&10.9777&14.1800\\
\hline
\hline
$\eta$&0.9680&0.9378 & 0.9908&0.9805\\
\hline
\end{tabular}
\label{Tab1}
\end{table}
\normalsize
Table \ref{Tab1} shows some optimal degree distributions for different $D$ and $\mu_o$, when $\epsilon=0.01$. As can be seen in this table, with increasing maximum degree $D$, the maximum efficiency of the code increases. This is due to larger degrees of freedom when increasing the maximum degree, which leads to a larger feasibility region for the linear program. Moreover, with increasing $\mu_o$, the maximum rate efficiency decreases, which is due to the larger number of inequality constraints in the linear program and accordingly smaller feasibility region.

\begin{figure}
  \centering
  \includegraphics[scale=0.30]{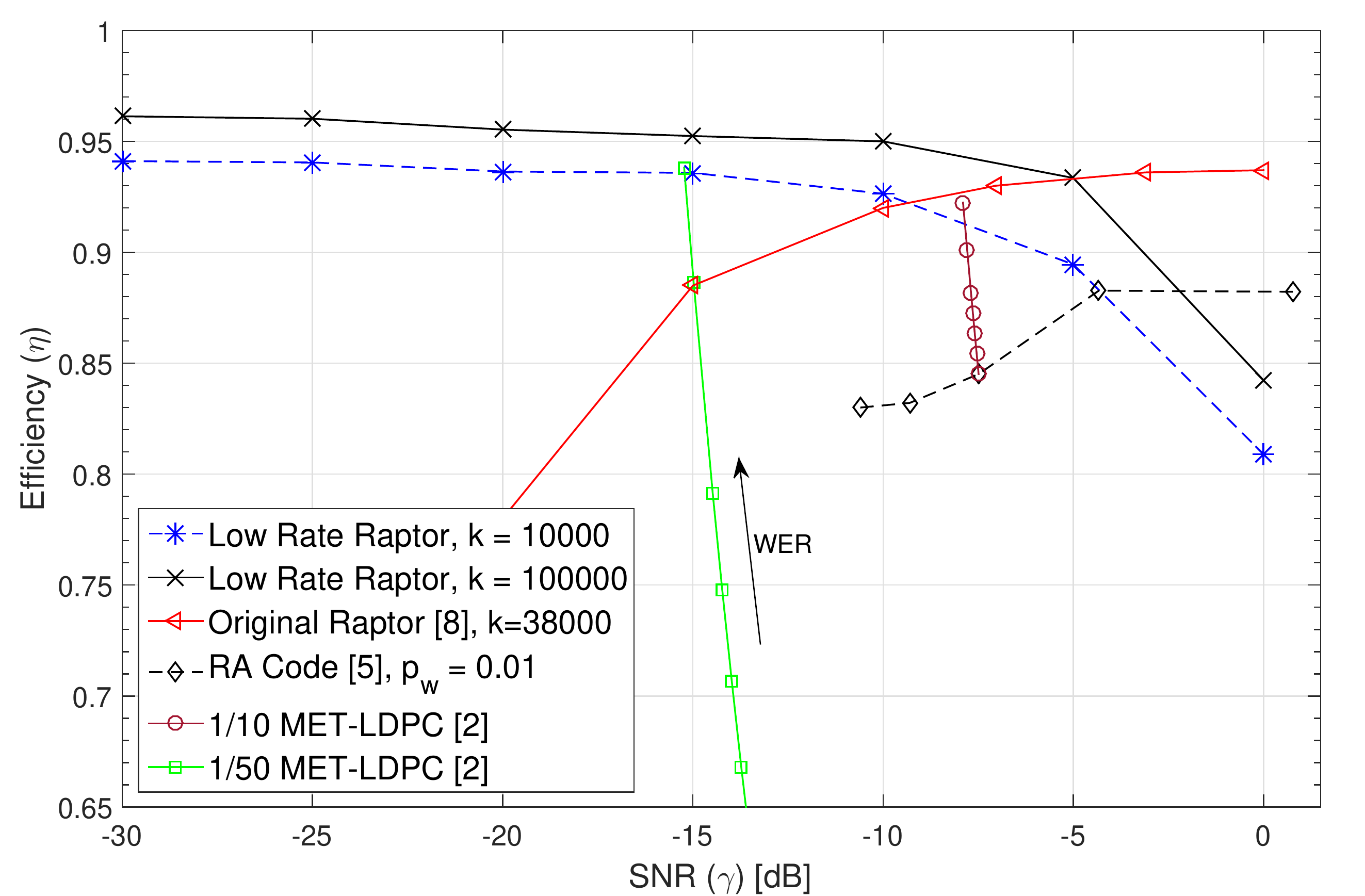}
  \caption{Rate efficiency versus the SNR for a Raptor code with the optimized degree distribution and different message lengths, when $D=300$. The code length for RA codes \cite{Sarah_J_RA_QKD} is 64800 and the code length for MET-LDPC is $10^{6}$ \cite{PhysRevA}. The WER for rate 1/50 and 1/10 MET-LDPC codes are respectively $\{0.54,0.16,0.038,0.021,0.014,0.012,0.009\}$ and $\{1.8\times10^{-3},5.7\times10^{-4},1.2\times10^{-4},3\times10^{-5},1.5\times10^{-5}\}$ from top to the bottom.}.
  \label{EffFig}
\end{figure}

Fig. \ref{EffFig} shows the rate efficiency versus the SNR for a Raptor code by using the optimized degree distribution listed in Table \ref{Tab1}, when the maximum degree is 300 and $\mu_o=40$. As can be seen in this figure, the code approaches efficiencies larger than 0.95 for very low SNRs when the message length is larger than 10000. Moreover, by increasing the message length, the rate efficiency improves which is due to the fact that the tree assumption and independence of the messages in the SPA decoding are more likely to be met as the message length increases. It is important to note that the efficiency at SNRs larger than -10 dB using the degree distributions obtained from Table \ref{Tab1} is not high enough which is due to the fact that these degree distributions were designed for very low SNRs. For SNR larger than -10 dB, one could solve the original linear program (\ref{optorig}) to find the optimal degree distribution, which has been shown to achieve efficiencies larger than 92\% at those SNRs \cite{Raptor_WideSNR}. The rate efficiencies for these codes have been also shown in Fig. \ref{EffFig}.

We have also compared rate 1/10 and 1/50 MET-LDPC codes and several RA codes in \cite{Sarah_J_RA_QKD} in Fig. \ref{EffFig}. As can be seen in this figure, Raptor codes with an appropriately designed degree distribution can achieve higher efficiencies in comparison with the fixed rate codes in the entire SNR range. It is important to note that LDPC and MET-LDPC codes can be designed for different rates, but, existing techniques fail to design very high efficient low rate codes for the low SNR regime. Moreover, the rate efficiency of fixed rate codes decreases with decreasing word error rate, which is also clear from Fig. \ref{EffFig}.

\section{Raptor Codes for CV-QKD Systems}
In the error reconciliation protocols with fixed rate LDPC codes, the receiver may not completely decode the key due to the non-zero word error rate of the code; thus throwing them away. This will limit the total key rate, as the total efficiency of the LDPC codes depends on the word error rate. This also has implications for security because throwing away erroneous codewords is a form of post selection. As the focus of this paper is on the coding efficiency of the proposed Raptor code over a CV-QKD systems, we omit the details of the Gaussian channel modeling and the reverse BI-AWGN channel in CV-QKD with reverse reconciliation due to space limitations. More details of this physical model  can be found in \cite{Fossier}.
\begin{figure}[!t]
  \centering
  \includegraphics[scale=0.38]{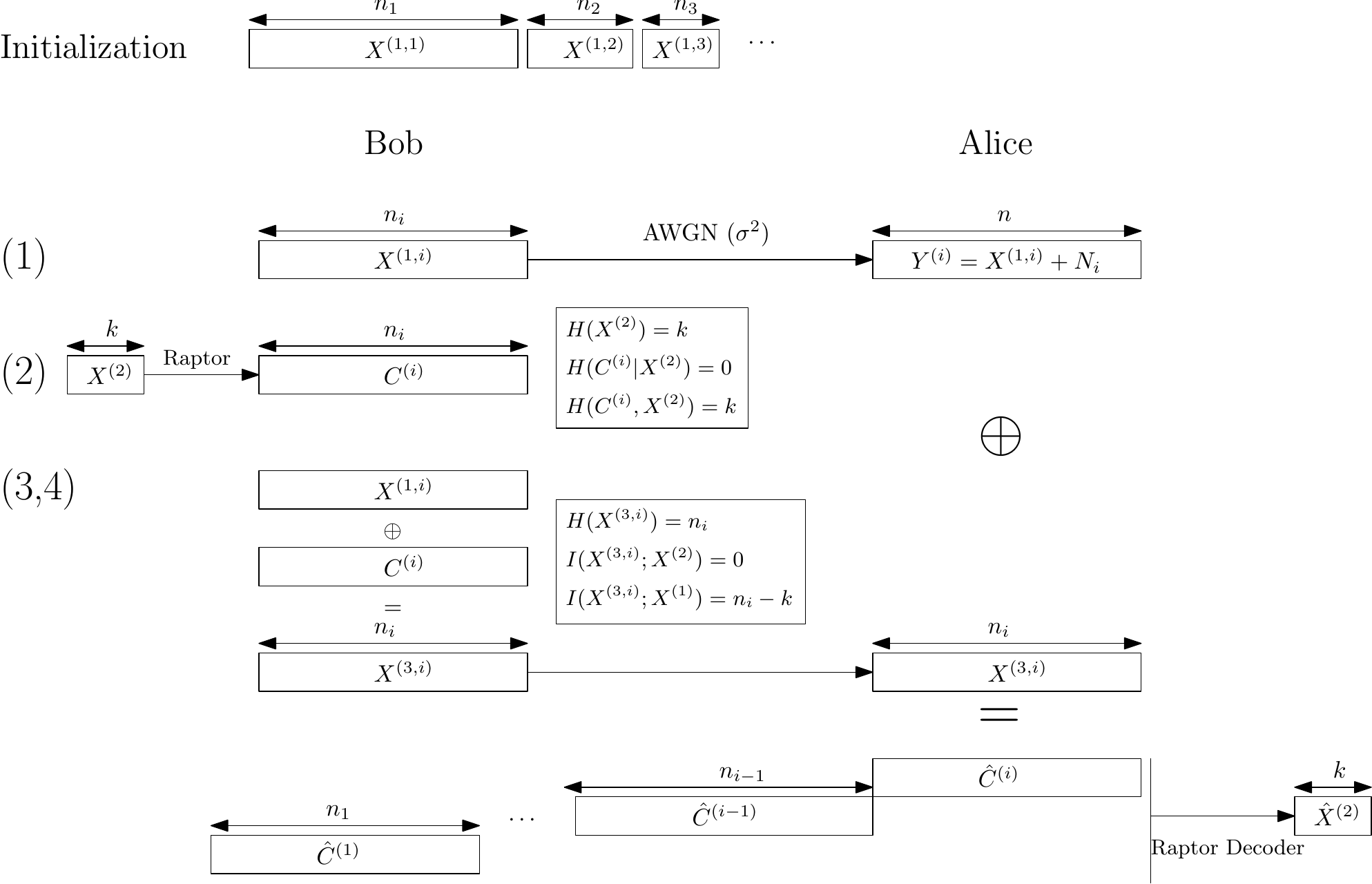}
  \caption{The proposed error reconciliation protocol using Raptor codes.}
  \label{p1}
\end{figure}

We propose to use Raptor codes to potentially send an unlimited number of coded symbols to the receiver, until it can successfully decode the key. No post selection is required as the WER is zero. This protocol is similar to \cite{PhysRevA}, but it runs in an iterative manner. Let $k$  denote the desired length of the key. The steps of the proposed CV-QKD protocol employing the reverse reconciliation are as follows. Note that we have excluded the steps regarding the quantum encoding, multidimensional mapping, and noise measurements due to space limitations (see \cite{PhysRevA} for further details).
\begin{itemize}
\item[$\mathrm{S}_1$] Bob encodes random bits $X^{(1)}$ onto quantum states and transmits them over a quantum channel to Alice who measures the received states $Y$.  Here $Y = X^{(1)}+ N$ where $N$ is additive white Gaussian noise with variance $\sigma^2$. Alice and Bob estimate the SNR (equivalently $\sigma$, where $\mathrm{SNR}=1/\sigma^2$). For simplicity we assume the transmissions used to estimate the noise are separate to the states in $X^{(1)}$.
\item[$\mathrm{S}_2$]  A $k$-bit key, $X^{(2)}$, is generated randomly. Bob then chooses a sequence of integers, $n_1$, $n_2$, ..., that are chosen such that $k/n_1=C(\sigma^2)$, $k/(n_1+n_2)=0.99 C(\sigma^2)$, and so on.
\end{itemize}
The remainder of the steps are iterated until a successful decoding has occurred. In the $i^{th}$ iteration, Alice and Bob do the following for $i\ge1$:
\begin{itemize}
\item[$\mathrm{S}_3$]  The $k$ bit message $X^{(2)}$ is encoded to a Raptor codeword $C^{(i)}$ of length $n_i$ using a Raptor code with the predetermined degree distribution function.
\item[$\mathrm{S}_4$] Using $n_i$ of the bits sent over the quantum channel the vector $X^{(3,i)}=X^{(1,i)}\oplus C^{(i)}$ is sent to Alice. Alice can then subtract $Y_i$ from $X^{(3,i)}$ to obtain $\hat{C}^{(i)}$. If $Y^{(i)}$ is error free $\hat{C}^{(i)}$ will be identical to $C^{(i)}$. A Raptor decoder is used to decode the key, where all received signals from previous iterations, i.e., $[\hat{C}^{(1)}, ..., \hat{C}^{(i-1)}, \hat{C}^{(i)}]$ are considered as the codeword. If decoding is successful, Alice sends an acknowledgment to Bob to stop transmitting new parity bits. Otherwise, step $\mathrm{S}_3$ and $\mathrm{S}_4$ are repeated with $i=i+1$.
\end{itemize}
Note that the key length is fixed, but the number of transmissions required on the quantum channel are not. Fig. \ref{p1} shows the steps of the error reconciliation in the proposed scheme using Raptor codes. The proposed scheme has the following advantages, which make it an excellent choice for CV-QKD systems.
\begin{itemize}
\item The encoding/decoding is linear in terms of the message length. This means that linear-time practical encoder/decoder can be implemented for them.
\item Unlike the design of fixed rate codes which is not always straightforward for all SNRs, a linear program can be defined to find the optimal degree distribution of a Raptor code for any SNR.
\item Raptor codes achieve very high efficiencies in very low SNR values, where the existing LDPC codes poorly perform in practice. This enables the CV-QKD systems to operate over very large distances.
\item Conventional fixed rate codes usually show poor performance in very low word error rates when the block length is finite. This significantly reduces the key rate of the CV-QKD systems. But with Raptor codes, an error free message is always delivered at the receiver by sending as many coded symbols as required by the receiver.
\end{itemize}
\section{Numerical Results}
We assume a coherent state CV-QKD system with homodyne detection and employing reverse reconciliation. The eavesdropper is assumed to utilize collective attacks. The CV-QKD model has the following free parameters. The signal variance encoded by Alice $V_A$, the channel loss $T$, the channel excess noise $\epsilon_n$, the homodyne efficiency $\eta_h$, and dark noise $v_{el}$ in Bob's station. The equivalent AWGN channel SNR can then be represented as follows \cite{PhysRevA}:
\begin{align}
\gamma=\frac{V_AT\eta}{2+\epsilon_n T\eta_h+2v_{el}}.
\end{align}
The maximum secret information available for an actual reconciliation with an imperfect code of efficiency $\eta$ is given by $\Delta I=\eta I_{AB}-I_{E}$, where $I_{AB}$ and $I_E$ are the mutual information of Alice and Bob and that of Eavesdropper, respectively. The secret key rate can be obtained as $\mathrm{Key Rate} = (1-p_w)(\eta I_{AB}-I_E),$ where $p_w$ is the WER of the code, which is zero for Raptor codes. An exact expression for the secret key rate of a CV-QKD system can be found in \cite{Fossier}. It is important to note that in this paper, we do not consider finite-size effects in the key rate calculation, which means that our numerical results show the secret key rate in the regime of infinite block length \cite{PhysRevA}. We also consider the loss in the code efficiency due to the multidimensional modulation, which has been studied in \cite{PhysRevA}. As shown in this paper, for an 8-dimensional reconciliation scheme, referred to as Octonion scheme,  the loss due to the multidimensional modulation is negligible in the low SNR regime.

\begin{figure}[!t]
  \centering
  \includegraphics[width=8cm, height=4.7cm]{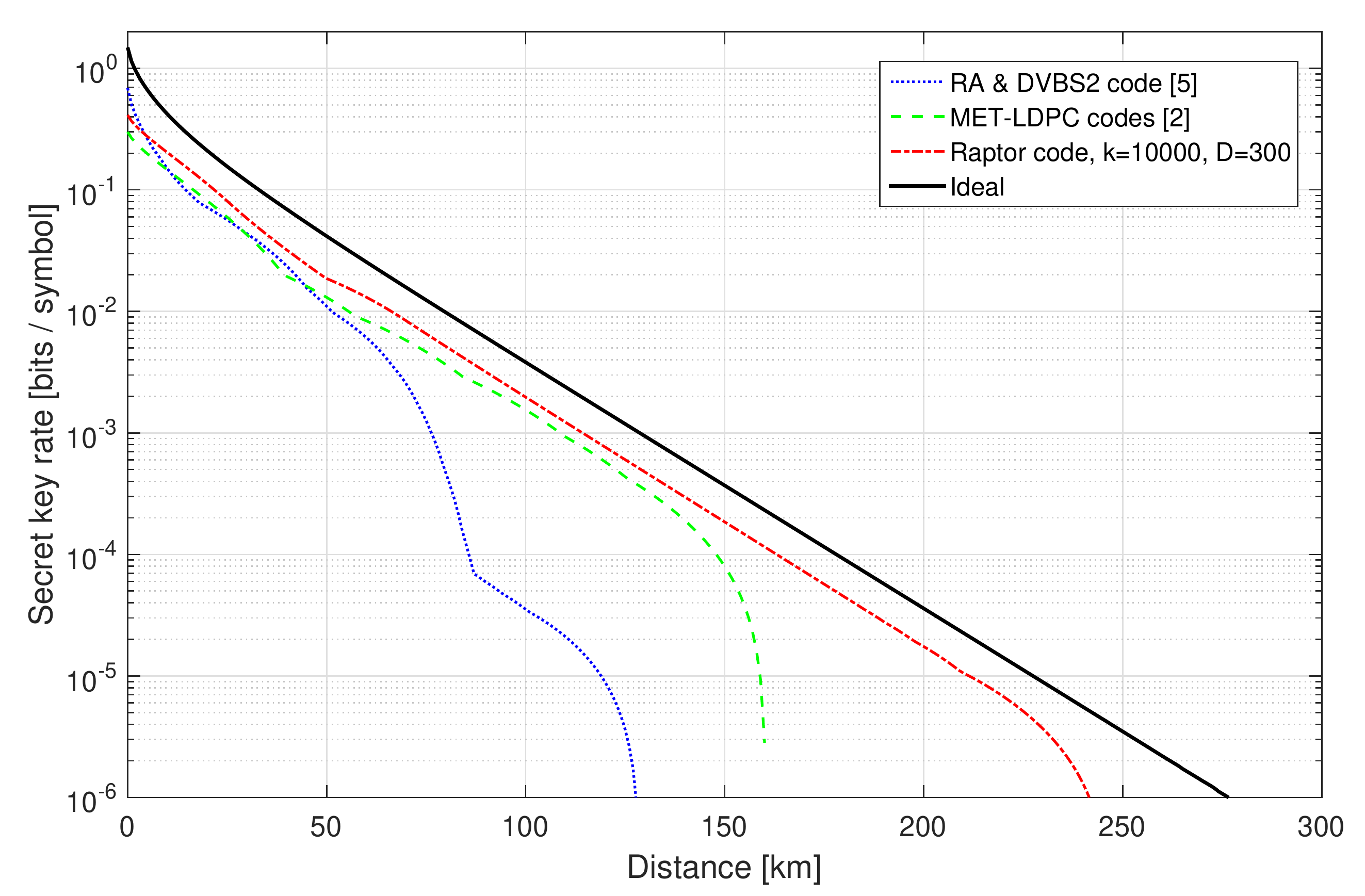}
  \caption{Secret key rate as a function of distance for a CV-QKD system with a homodyne detector and Raptor/RA/MET-LDPC codes. The code efficiencies for ME-LDPC and DVB-S2/RA codes are respectively taken from \cite{PhysRevA} for $p_w=1/3$ and from \cite{Sarah_J_RA_QKD} for  $p_w\in\{0.1,0.01\}$. $\eta_h=0.6$, $V_{el}=0.01$, and $\epsilon_n=0.01$.}
  \label{KeyrateCompFig}
\end{figure}

Fig. \ref{KeyrateCompFig} shows the optimized secret key rate as a function of the operational distance for a CV-QKD system with homodyne detection and different coding schemes when an attenuation of 0.2 dB/km is assumed. Note that DVBS2, RA and MET-LDPC codes show different rate efficiencies for different WER values. The results in Fig. \ref{KeyrateCompFig} are based on simulation results for finite length RA and ME codes with secret key optimization for these codes carried out over different SNRs and WERs. As can be seen in this figure, the proposed error reconciliation protocol with Raptor codes significantly outperforms the existing protocols over all distances. This comes from the fact that existing fixed rate codes shows very poor efficiencies ($< 75\%$) in very low SNRs even with a large WER (1/10), while Raptor codes achieve very high efficiencies ($> 95\%$) with zero WER in SNRs as low as -30 dB.


\section{Conclusions}
In this paper, we studied the design of degree distribution of Raptor codes in the low SNR regime and proposed a linear program to obtain the optimal degree distribution. Simulation results showed that Raptor codes achieve very high efficiencies in very low SNRs that makes them very attractive to be used in CV-QKD systems. We incorporated Raptor codes in the error reconciliation protocol for a CV-QKD system, and showed through numerical analysis that the proposed scheme achieves near optimal key rates and significantly outperformed the existing protocols based on fixed rate codes.

\bibliographystyle{IEEEtran}
\footnotesize
\bibliography{IEEEabrv,sample2}

\begin{thebibliography}{10}
\providecommand{\url}[1]{#1}
\csname url@samestyle\endcsname
\providecommand{\newblock}{\relax}
\providecommand{\bibinfo}[2]{#2}
\providecommand{\BIBentrySTDinterwordspacing}{\spaceskip=0pt\relax}
\providecommand{\BIBentryALTinterwordstretchfactor}{4}
\providecommand{\BIBentryALTinterwordspacing}{\spaceskip=\fontdimen2\font plus
\BIBentryALTinterwordstretchfactor\fontdimen3\font minus
  \fontdimen4\font\relax}
\providecommand{\BIBforeignlanguage}[2]{{%
\expandafter\ifx\csname l@#1\endcsname\relax
\typeout{** WARNING: IEEEtran.bst: No hyphenation pattern has been}%
\typeout{** loaded for the language `#1'. Using the pattern for}%
\typeout{** the default language instead.}%
\else
\language=\csname l@#1\endcsname
\fi
#2}}
\providecommand{\BIBdecl}{\relax}
\BIBdecl

\bibitem{PhysRevLett}
M.~Navascu\'es, F.~Grosshans, and A.~Ac\'{i}n, ``Optimality of {G}aussian
  attacks in continuous-variable quantum cryptography,'' \emph{Phys. Rev.
  Lett.}, vol.~97, p. 190502, Nov 2006.

\bibitem{PhysRevA}
P.~Jouguet, S.~Kunz-Jacques, and A.~Leverrier, ``Long-distance
  continuous-variable quantum key distribution with a {G}aussian modulation,''
  \emph{Phys. Rev. A}, vol.~84, p. 062317, Dec 2011.

\bibitem{PhysRevAD11}
A.~Leverrier and P.~Grangier, ``Continuous-variable quantum-key-distribution
  protocols with a non-{G}aussian modulation,'' \emph{Phys. Rev. A}, vol.~83,
  p. 042312, Apr 2011.

\bibitem{PhysRevAC11}
A.~Leverrier, R.~All\'eaume, J.~Boutros, G.~Z\'emor, and P.~Grangier,
  ``Multidimensional reconciliation for a continuous-variable quantum key
  distribution,'' \emph{Phys. Rev. A}, vol.~77, p. 042325, Apr 2008.

\bibitem{Sarah_J_RA_QKD}
\BIBentryALTinterwordspacing
S.~J. {Johnson}, V.~A. {Chandrasetty}, and A.~M. {Lance}, ``{Repeat-Accumulate
  Codes for Reconciliation in Continuous Variable Quantum Key Distribution},''
  \emph{ArXiv e-prints}, 2015. [Online]. Available:
  \url{http://arxiv.org/abs/1510.03510}
\BIBentrySTDinterwordspacing

\bibitem{Luby}
M.~Luby, ``{LT} codes,'' in \emph{Proc. 43rd Annual IEEE Symp. Foundations
  Comput. Science}, Nov. 2002, pp. 271 -- 280.

\bibitem{RaptorBSC}
O.~Etesami and A.~Shokrollahi, ``Raptor codes on binary memoryless symmetric
  channels,'' \emph{{IEEE} Trans. Inf. Theory}, vol.~52, no.~5, pp. 2033 --
  2051, May. 2006.

\bibitem{Raptor_BIAWGN}
Z.~Cheng, J.~Castura, and Y.~Mao, ``On the design of {R}aptor codes for
  binary-input {G}aussian channels,'' \emph{{IEEE} Trans. Commun.}, vol.~57,
  no.~11, pp. 3269--3277, Nov. 2009.

\bibitem{Raptor_WideSNR}
S.-H. Kuo, Y.~L. Guan, S.-K. Lee, and M.-C. Lin, ``A design of physical-layer
  {R}aptor codes for wide {SNR} ranges,'' \emph{{IEEE} Commun. Lett.}, vol.~18,
  no.~3, pp. 491--494, Mar. 2014.

\bibitem{Raptor}
A.~Shokrollahi, ``Raptor codes,'' \emph{{IEEE} Trans. Inf. Theory}, vol.~52,
  no.~6, pp. 2551 --2567, Jun. 2006.

\bibitem{Mathhandbook}
M.~Abramowitz and I.~A. Stegun, \emph{Handbook of mathematical functions: with
  formulas, graphs, and mathematical tables}.\hskip 1em plus 0.5em minus
  0.4em\relax Courier Corporation, 1964, no.~55.

\bibitem{Fossier}
S.~{Fossier}, E.~{Diamanti}, T.~{Debuisschert}, R.~{Tualle-Brouri}, and
  P.~{Grangier}, ``{Improvement of continuous-variable quantum key distribution
  systems by using optical preamplifiers},'' \emph{Journal of Physics B Atomic
  Molecular Physics}, vol.~42, no.~11, p. 114014, 2009.

\end{thebibliography}

\end{document}